\documentclass[aps,pra, twocolumn,showpacs,showkeys]{revtex4}

\usepackage{graphicx}
\usepackage{subfigure}

\newcommand{\ket}[1]{|#1\rangle}

\newcommand{\expt}[1]{\langle #1 \rangle}
\newcommand{\Tr}[1]{\mathrm{Tr} \left({#1}\right)}

\begin{document}

\title{Weak measurement and control of entanglement generation}

\author{Charles Hill}
\email{Charles.Hill@liverpool.ac.uk} \affiliation{Department of
Electrical Engineering and Electronics, University of Liverpool,
Brownlow Hill, Liverpool, L69 3GJ, United Kingdom}

\author{Jason Ralph}
\email{jfralph@liverpool.ac.uk} \affiliation{Department of Electrical
Engineering and Electronics, University of Liverpool, Brownlow Hill, Liverpool,
L69 3GJ, United Kingdom}




\begin{abstract}
In this paper we show how weak measurement and local feedback can be
used to control entanglement generation between two qubits. To do
this, we make use of a decoherence free subspace (DFS). Weak
measurement and feedback can be used to drive the system into this
subspace rapidly. Once within the subspace, feedback can generate
entanglement rapidly, or turn off entanglement generation
dynamically. We also consider, in the context of weak measurement,
some of differences between purification and generating
entanglement.
\end{abstract}

\maketitle


Entanglement is a fundamental resource used in many quantum
information applications  \cite{NC01}. This paper considers how weak
measurements and local feedback can be used to control the
generation of entanglement between two qubits. There are three main
points that this paper makes. First, it is possible to speed up the
generation of entanglement using feedback. Second, it is possible to
start and stop entanglement production. Third, the protocol for
increasing the rate purification is not necessarily the same as
increasing the rate of entanglement.

Control of entanglement generation can be achieved with the aid of a
decoherence free subspace (DFS) \cite{Bac00, Dua97, Lid98, Zan97} in
which the measurement no longer produces any useful information
about the state of the quantum system - a measurement free subspace.
Weak measurement and feedback is used to (i) force the system into
the subspace, and (ii) to purify the system using the structure of
the subspace. The technique proposed here is analogous of existing
single qubit rapid state reduction protocols in each of the
subspaces defined by the measurement \cite{Jac03, Com06, Wis06}. The
second process is particularly interesting because it takes the
system from a classically correlated state to a maximally entangled
Bell state, generating entanglement more rapidly than would be
achieved without feedback. Using the structure of the DFS,
entanglement generation may also be switched on an off dynamically,
without turning off the measurement interaction (an important factor
in many solid state qubits). This control may be achieved by
applying local Hamiltonian feedback alone. Together local unitary
control and weak measurement allow rapid generation of entanglement
and control of the level of entanglement which is ultimately
generated.


Weak measurement makes it possible to modify the evolution
continuously via \emph{Hamiltonian feedback}, where the Hamiltonian
feedback applied to the system depends on the measurement record
\cite{Bel87, Kor99, Oxt05, Wis93, Hof98, Sto04, Wis94, Doh99, Com06,
Wis07, Jor06, Ple99}. Hamiltonian feedback during measurement not
only affects the final state of the system, but also the measurement
process itself. For example it can affect the rate of state
reduction/purification. In a protocol described by Jacobs for a
single qubit, the average rate of state reduction (as measured by
the purity) for a single qubit can be maximized by feedback
\cite{Jac03}. This process is known as rapid state reduction
\cite{Com06}, or as rapid purification \cite{Jac03}. Jacobs'
protocol is deterministic, but other protocols exist which are
stochastic and minimize the average time for a single qubit to reach
a given purity \cite{Wis06}. Combes and Jacobs demonstrated that
similar feedback can be applied to higher dimensional systems to
increase the average increase in purity \cite{Com06}. More recently
Wiseman has proved that these protocols are optimal \cite{Wis07}.
These methods have also been applied to bipartite systems
\cite{Hil07}.

In this paper we consider two continuously monitored qubits. In
contrast to previous  protocols which aim to increase the rate of
purification of the system, \cite{Com06, Hil07} we concentrate on
speeding up the generation of entanglement between the two qubits.
These two goals are compatible. It is possible to view the
entanglement generating technique as a modification of the
previously existing purification protocols \cite{Jac03, Com06},
applied to an logical qubit.

Generation of entanglement via measurement should also be contrasted
with generation through Hamiltonian control alone
\cite{Duer01,Kra01}. In this paper the weak  measurement is
responsible for entanglement generation between the systems which
are not directly interacting via an entangling Hamiltonian, and the
control presented here is closed loop, rather than open loop.
%



Weak measurement is modeled by a stochastic master equation (SME).
The SME is obtained by introducing an ancilla system, weakly coupled
to the system of interest. The auxiliary system then undergoes a
measurement giving a stochastic result, and is then traced out. This
leaves only the system of interest, $\rho$, and a stochastic
measurement record $r(t)$ which can be used to construct an estimate
of the state of the system, $\rho$. This is often referred to as an
`unraveling' of the master equation\cite{Bel87, Cav87, Car89,
Kor99,Oxt05}.

The stochastic master equation (SME) which governs the evolution of
the density operator $\rho$ in the presence of a weak measurement of
a Hermitian observable, $y$, in the absence of a driving Hamiltonian
is given by
\begin{equation}
d\rho = -k[y,[y,\rho]] dt + \sqrt{2k}(y \rho + \rho y - 2 \langle y
\rangle \rho) dW, \label{eqn:SME}
\end{equation}
where  $k$ is the measurement strength. The first term in this
equation describes the familiar drift towards the measurement axis.
The second term in the equation is weighted by $dW$, a Wiener
increment with $dW^2 = dt$. This term describes the update of
knowledge of the density matrix conditioned on the measurement
record \cite{Doh99}.

We consider a specific case: the observable, $\hat{y}=ZZ$, which is
continuously monitored on the  system. $Z \ (X, Y)$ is short hand
for the Pauli matrix, $\sigma_3 \ (\sigma_1, \sigma_2)$ and the
tensor product is implied (ie $ZZ=Z\otimes Z$). This parity
measurement may arise when the measurement device is coupled to both
qubits \cite{Tra06, Mao04, Rus03, Bee04}. Theoretically it is
arguably the simplest measurement which can be performed on the
system, only giving information about whether the spins are aligned
or anti-aligned. It is also the type of measurement envisioned for
error-correction codes \cite{Ahn02}. The measurement record,
\begin{equation}
dr(t) = \expt{y} dt + \frac{dW}{\sqrt{8k}}
\end{equation}
will depend not on the reduced state of one qubit or the other, but
on the correlations between the two qubits.

In this paper we limit ourselves to applying only local Hamiltonian
feedback to one qubit or the other. This is a purely practical
constraint; many quantum systems exhibit local unitary control.
%

As a measurement of $y=ZZ$ is made, the system evolves according to
the stochastic master equation:
\begin{eqnarray}
d\rho &=& -k[ZZ,[ZZ,\rho]] dt \nonumber \\
&& + \sqrt{2k}(ZZ \rho + \rho ZZ - 2 \langle ZZ \rangle \rho) \ dW.
\label{eqn:smeZZ}
\end{eqnarray}
The density matrix can be expanded in the Pauli basis
\begin{equation}
\rho = \sum_{i,j = X,Y,Z,I} \frac{r_{ij}}{d} \sigma_i \sigma_j.
\end{equation}
In which $d$ is the dimension of the system (that is $d=4$) and the
coefficients, $r_{ij}$, may be found by $r_{ij} = \Tr{\sigma_i
\sigma_j \ \rho}$, which should not be confused with the measurement
record, $r(t)$.

Previous schemes using weak measurement and feedback have often
concentrated on the purity of the system. The purity of the system
is given by $P(\rho) = \Tr{\rho^2}$. A completely pure state has a
purity of $1$, and a completely mixed state has a purity of $1/d$.
For a state which is confined to a particular subspace of the
system, but completely mixed within that subspace, the purity is
given by $1/d_s$, where $d_s$ is the dimension of the subspace.

In the case of a measurement of $ZZ$ the total purity of the system evolves
according to the equation,
\begin{eqnarray}
dP&=& 8 k \big( \sum_{ij} (r_{ZZ \sigma_i\sigma_j} - r_{ZZ} r_{ij})^2 - \sum_{mn} (1-r_{ZZ}^2)r_{mn}^2 \big) dt \nonumber \\
&+& 4 \sqrt{2k} \big( \sum_{ij} (r_{ZZ \sigma_i\sigma_j}  - r_{ZZ}
r_{ij}) r_{ij} - \sum_{mn} r_{mn}^2 r_{ZZ} \big) dW.
\end{eqnarray}
where $m$ and $n$ range over all the Pauli matrices which anti-commute with
$ZZ$, and $i$ and $j$ range over all the Pauli matrices which commute with $ZZ$.

%

In order to quantify the correlations between the qubits, we will
use the value $R_2^2 = \sum_{i,j = X,Y,Z} r_{ij}^2$. $R_2^2$ has a
maximum value of $3$ when the system is in a maximally entangled
state (a pure Bell state). $R_2^2$ has a minimum of $0$ when there
are no classical correlations between the states of the two qubits.
$R_2^2 \le 1$ for a product state. For a mixed state, increasing
$R_2^2$ to its maximum value leads to an increase in both purity and
entanglement in the qubits. $R_2^2$ is invariant under single qubit
rotations of either system. It is similar to purity, for which rapid
purification protocols are available. It is simple to calculate,
monitor and to obtain analytic results for.




%


We now show how this two-qubit SME can be applied to decoherence
free subspaces (DFS). DFS were introduced as a way to protect
fragile quantum information passively from the effects of an
interaction with an environment \cite{Bac00,Dua97,Lid98,Zan97}.
Under simple assumptions about symmetries of the coupling between
the system and the environment, there are subspaces of the overall
Hilbert space which remain unaffected by the interaction of the
system and its environment. Therefore any information encoded in
this subspace is protected from decoherence.

We now consider the system studied in this paper, where the
interaction is given by $ y = Z\otimes Z$. In this case, there are
clearly two degenerate eigenvalues, $+1$ and $-1$. The corresponding
measurement free subspaces are given by $ D_{+} = \mathrm{Span} \{
\ket{00}, \ket{11} \}$, and $D_{-} = \mathrm{Span} \{ \ket{01},
\ket{10} \}$.

Consider the stochastic master equation (\ref{eqn:SME}) acting on a
state, $\rho$, restricted to a DFS (either $D_+$ or $D_-$). A DFS is
found by finding degenerate eigenstates of error generators,
$F_\alpha$, as well as a measurement, $y$. Then by definition each
of the basis vectors of the DFS are degenerate eigenstates of $y$:
$y \ket{i} = c_y \ket{i}$. This means that $[y, \rho]=0$ and that
for states restricted to the DFS $\expt{y} = \mathrm{Tr}[y\rho] =
c_y$. Therefore if $\rho$ is restricted to a DFS whose error
generators include the measurement operator, $d\rho = 0$. There is
no change in the conditional density matrix $\rho$ according to the
stochastic master equation.




%

Once restricted to a measurement free subspace, further measurement
of $y=ZZ$ yields no useful information about the system. However, by
applying local rotations to the system, it may be rotated outside
the DFS where measurement provides further useful information. If
Hadamard (H) gates (a local rotation by $\pi$ around $(X + Z) /
\sqrt{2}$) are applied locally to each of the two qubits then the
system will be rotated out of the DFS. Since $H\otimes H \ Z\otimes
Z \ H\otimes H = X\otimes X$ commutes with $Z\otimes Z$, we will
always be able to rotate back to the original subspace. Therefore it
is possible to turn the measurement of the system on and off by
rotating it into and out of the DFS, even when the measurement
apparatus is always interacting with the system. This can be
performed using local rotations alone - Hadamard is a local unitary
operation.

Often the goal of measurement is to prepare an experimental system
in a given state, for example: to prepare states of a given purity.
The protocol described here could be applied to a large ensemble of
qubit pairs. If the purity were monitored continuously, then the
measurement could be turned off dynamically once it reached the
desired value (the time of first passage) \cite{Wis06}. Eventually
all the qubits will have reached the threshold, and would be held in
their DFS. This would allow a large number of systems to be prepared
simultaneously.


We now show how the system can be viewed as two encoded qubits. Each
encoded qubit has a state whose evolution depends on the collective
behaviour of both physical qubits. The first encoded qubit
represents the extent to which information is found within one
subspace, $D_+$, or the other $D_-$. The second encoded qubit
represents the information protected within the DFS. We will then
adapt Jacobs' protocol for a single qubit to apply not to physical
qubits, but to the encoded qubits.


%
%

First, we will describe in more detail how the system can be viewed
as two encoded qubits. There is a natural way to divide the space to
represent an encoded qubit. Elements of the commutant of $ZZ$ (eg.
$X\otimes X$) leave the system inside the DFS (eg. $D_+$). Therefore
they can be considered as operating on an encoded qubit which is
immune from measurement of $y=ZZ$.

Similarly physical operations which do not commute with $ZZ$ can be
considered as rotations of a separate encoded qubit. For example,
$X\otimes Z$ could be considered an encoded operation applying an
$x$-rotation to the first of the two encoded qubits. In fact, it is
possible to construct an entire encoded Bloch sphere. Each
\begin{equation}
 H_{ZZ} = Z_L \otimes I_L, \  H_{XZ} = X_L \otimes I_L, \ H_{YI} = Y_L \otimes I_L,
\end{equation}
can be considered as designating an encoded Bloch sphere on the
first encoded qubit. This qubit represents the extent to which the
the system is confined to the DFS, with the states $\ket{0_L}$ and
$\ket{1_L}$ representing states which are entirely confined to one
of the spaces or the other.

We apply Jacobs' single qubit protocol to the \emph{encoded} qubit
\cite{Jac03}. Jacobs showed that the fastest average rate of
purification was obtained when the Bloch vector was rotated away
from the measurement axis. In this case,  the encoded qubit is
confined to the $X\otimes Z$ axis, by measuring $Z\otimes Z$ and
applying local feedback of rotations around the $Y\otimes I$ axis.
That is, feedback is only applied locally, to the first of the two
physical qubits.

By applying Hamiltonian feedback it is possible to continually
rotate the state of the encoded qubit so that it lies along the
$X\otimes Z$ axis. In this case, the average rate of purification is
largest on average for the encoded qubit. Purifying the first qubit
is equivalent (up to a local rotation) to confining the system to a
decoherence free subspace. So rapidly purifying the state of the
first encoded qubit rapidly forces the system into the DFS.

Another advantage of Jacobs' protocol is that the rate of
purification is deterministic. If the feedback is perfectly applied,
the purity of the system can be predicted. A large number of systems
undergoing purification to the DFS would all be purified at the same
rate, and would achieve the same level of purity at the same time.
Once the system has been purified to the DFS, it is in a classically
correlated state. Further purification takes the state to a
maximally entangled Bell state. This process is achieved more
rapidly with feedback than without.

The states described in this section represent the information
encoded within the DFS. In order to operate of the state of a
protected qubit, we choose a separate set of encoded operations:
\begin{equation}
K_{XX} = I_L\otimes X_L, \ K_{XY} = I_L\otimes Y_L, \ K_{IZ} =
I_L\otimes Z_L.
\end{equation}
Here each rotation is chosen from the commutant, $A'$. In other words, every
operation commutes with $y$.

If it was possible to measure $X\otimes X$ directly, and apply
Hamiltonian feedaback, $I\otimes Z$, to the second physical qubit
then we would be able to apply Jacobs' protocol to rotate the state
onto the $X\otimes Y$ axis. Seen in the encoded space, this is
simply another application of Jacobs' rapid purification method
applied to the second encoded qubit. Measurement of $X\otimes X$ is
equivalent to Hadamard gates applied in parallel to both physical
qubits, followed by measurement of $Z\otimes Z$. Local Hamiltonian
feedback can be applied by performing $x$-rotations on the second
physical qubit to rotate the state the $ZY$ axis. This scheme
requires only local unitary operations, and measurement of $Z\otimes
Z$.

\begin{figure}
\begin{center}
\includegraphics[width=\columnwidth]{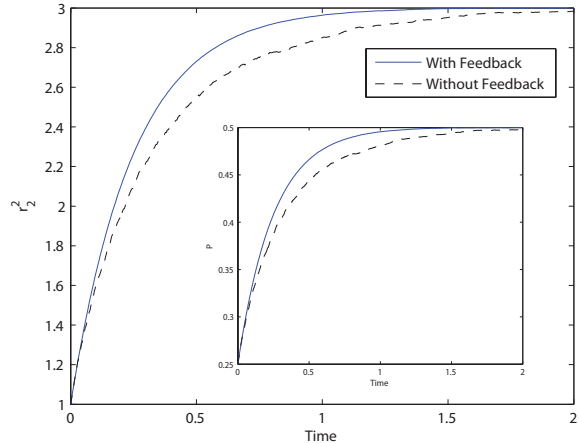}
\caption{The average correlation (over 1000 runs) demonstrating
rapid entanglement, with and without feedback. The inset shows the
purity of this system as it is brought into to this subspace.}
\label{fig:InsideDFSPurity}
\end{center}
\end{figure}

This is shown in Figure \ref{fig:InsideDFSPurity}. In this case, we are going
from a classically correlated state in the DFS to a maximally entangled Bell
state, and so this process could be viewed as rapid entanglement of the two
qubits. It is evident that the average rate of increase in purity
with feedback is faster than without feedback.

The protocols which provide the fastest rate of increase of purity
are not necessarily the same as those that provide the fastest rate
of increase of entanglement. This is most easily seen by considering
the two encoded qubits. Consider the simplest case, when the encoded
qubits are separable, and the second encoded qubits is not in a pure
state. In this case, to increase the purity of the system a
measurement should be made to the second encoded qubit. However,
this does not necessarily lead to an entangled state. In particular
if the first encoded qubit is in the state $\ket{0_L}$ (that is
$\expt{Z\otimes Z}=1$), and the second encoded qubit is being
purified along the $I\otimes Z$ axis, then this scheme gives the
greatest increase in purity, but the entanglement of the system
stays the same (ie. 0). Conversely if the second qubit is not quite
pure, but has $\expt{X\otimes X} \approx 1$ and the first qubit has
$\expt{Y\otimes I}=1$ then measurements on the first encoded qubit
will not change the purity of the system, however they can be used
(as described in this paper) to force the two physical qubits
towards an entangled state such as $\expt{Z\otimes Z}=\pm 1$. In
that case, the entanglement of the system increases, but the purity
remains the same. Therefore the protocols which should be applied to
increase the level of entanglement in a system are not necessarily
the same as those which have been proposed to increase the level of
purity.

In this paper we have considered how it is possible to use local
measurement and an always-on measurement to control the entanglement
of a bipartite quantum system. First we considered if it is possible
to turn off the entangling effect of measurement using local
rotations alone. We also showed that it is possible to start and
stop entanglement production, using well known decoherence free
subspaces. We showed how it was possible to guide the system into
the decoherence free subspace using feedback. With feedback this
operation proceeds faster than is achieved without feedback.

Once in the decoherence free subspace, we showed that it is possible
to speed up the generation of entanglement using feedback. We
achieved this by taking the system from a classically correlated
state to a maximally entangled Bell state. This process proceeds
faster, on average, when it is possible to apply local qubit
feedback based on the measurement record. The feedback required is
simple. It depends on only two parameters, and can be applied using
local Hamiltonians alone. We refer to this process as rapid
entanglement. Both of these process could be seen as applying
existing purification protocols to encoded qubits, rather than
physical qubits. However, as we showed, purification is quite
different from generating entanglement. In particular it is possible
to speed up the generation of entanglement (whether using feedback
or not) without purifying the system.

The authors would like to acknowledge that this work was supported
by UK EPSRC grant number EP/C012674/1.

\bibliography{bibliographyLiverpool}

\end{document}